\documentstyle{article}
\font\Sets=msbm10
\def\Real{\hbox{\Sets R}}
\def\bb{\begin{equation}}
\def\ee{\end{equation}}

\newtheorem{theorem}{Theorem}

\begin{document}
\begin{center}
{\large\bf
The Fourier method for the linearized Davey-Stewartson I equation
\\
\vspace*{3mm}
O. M. Kiselev}
\\ \vspace*{3mm}
Ufa Institute of Mathematics, Russian Academy of Sciences,\\
112 Chernyshevsky str., Ufa, 450000, Russia\\
E-mail: ok@imat.rb.ru
\end{center}
\begin{abstract}
{The linearized Davey-Stewartson equation with varing coefficients
is solved by Fourier method. The approach uses the inverse scattering
transform for the Davey-Stewartson equation.}
\end{abstract}

\section{Introduction}
\par
The Davey-Stewartson equation (DS) is the well-known subject for
in\-ves\-ti\-ga\-tions because of two causes.
First, the equation models a nonlinear
interaction between a long surface wave and a short surface wave \cite{D-S}.
Second, DS is integrable by an inverse scattering transform (IST)
\cite{F-A} and then one can investigate a solution structure in details.
\par
In this work we consider a linear system of equations with varing
coefficients, which come out when one linearizes the DS at a nonzero
solution as a background. We develope the Fourier method for the above
linear system. This method is based on IST results for DS \cite{F-A,F-S,N}.
The basic functions, which are used for the Fourier expansion, are
assosiated with a scattering problem for a Dirac system.
\par
A realized approach one can see as generalizing on (2+1)-dimensions
(two spatial variables and time) system of pioneer works of D.Kaup
\cite{K1,K2}, in which the Fourier method was formulated for linearized
(1+1)-dimensional integrable equations. The obtained results make
possible to study perturbations of the DS, in general putting out
the equation from the class of integrable equations.

\section{The Dirac equation and the basic functions}
\par
Here we consider a Goursat problem for the Dirac equation
\cite{F-A}-\cite{N}:
\bb
\Bigg(
\begin{array}{cc}
\partial_\xi & 0\\
0 & \partial_\eta
\end{array}\Bigg)
\psi=-{1\over2}\Bigg(
\begin{array}{cc}
0 & q_1\\
q_2 & 0
\end{array}\Bigg)\psi
\label{de}
\ee
Define a solutions of (\ref{de}) as follows \cite{F-A}:
$$
\begin{array}{cc}
\psi^+_{11}|_{\xi\to-\infty}=\exp(ik\eta),&\psi^+_{12}|_{\xi\to-\infty}=0,\\
\psi^+_{21}|_{\eta\to\infty}=0,&\psi^+_{22}|_{\eta\to-\infty}=\exp(-ik\xi);\\
\psi^-_{11}|_{\xi\to-\infty}=\exp(ik\eta),&\psi^-_{12}|_{\xi\to+\infty}=0,\\
\psi^-_{21}|_{\eta\to-\infty}=0,&\psi^-_{22}|_{\eta\to-\infty}=\exp(-ik\xi).\\
\end{array}
$$
\par
Denote by $(\chi,\mu)_{q}$ a bilinear form:
\bb
(\chi,\mu)_q=\int_{-\infty}^{\infty}\int_{-\infty}^{\infty}d\xi d\eta
(\chi_1\mu_1\,q_2\,+\,\chi_2\mu_2\,q_1),
\label{f1}
\ee
where $\chi_i$ and $\mu_i$ are elements of the columns $\chi$ and $\mu$.
\par
Denote by $\phi^{(1)}$ and $\phi^{(2)}$ the solutions which are dual
to $\psi_{(1)}$ (first column of the matrix $\psi^+$) and
$\psi_{(2)}$ (second column of the matrix $\psi^-$) with respect to
the bilinear form (\ref{f1}).
\par
It is known, that the solution of (\ref{de}) satisfies a nonlocal
Riemann-Hilbert problem \cite{F-S}. Formulate this problem for
$\psi^-_{11}$ and  $\psi^+_{12}$. Denote by
$\psi^{(1)}$ a row $\{\psi^-_{11},\psi^+_{12}\}$, then \cite{F-S}:
$$
\psi^{(1)}=E^{(1)}(ik\eta)+S[s]\psi^{(1)}.
$$
Here $E^{(1)}(z)$ is the first row  of a matrix
$E(z)=diag(\exp(z),\exp(-z))$, the operator  $S[s]$ is defined by formula:
$$
S[s]\psi^{(1)}=\big[\exp(ik\eta)\bigg(\exp(-ik\eta)\int_{-\infty}^{\infty}dl
s_1(k,l)\psi^+_{12}(\xi,\eta,l)\bigg)^-,
$$
$$
\exp(ik\xi)\bigg(\exp(-ik\xi)\int_{-\infty}^{\infty}dl
s_1(k,l)\psi^-_{11}(\xi,\eta,l)\bigg)^+ \big],
$$
where
$$
\bigg(f(k)\bigg)^{\pm}={1\over2i\pi}\int_{-\infty}^{\infty}
{dk'\,f(k')\over k'-(k\pm i0)}.
$$
\par
Denote by $<\chi,\mu>_s$ a bilinear form:
\bb
<\chi,\mu>_s=\int_{-\infty}^{\infty}\int_{-\infty}^{\infty}dk dl
(\chi^1(l)\mu^1(k)\,s_2(k,l)\,+\,\chi^2(l)\mu^2(k)\,s_1(k,l)),
\label{f2}
\ee
where $\chi^{j}$ is the element of the row $\chi$.
\par
Let  $\varphi^{(j)},\, j=1,2$ be rows which are solution
of the equations conjugated to the equation for the rows
$\psi^{(j)}=\{\psi^-_{j1}, \psi^+_{j2}\}$ with respect to the
bilinear form (\ref{f2}).
\begin{theorem} Let  $q_1$ and $q_2$ satisfy following conditions:
$\partial^{\alpha} q_{1,2}\in L_1\cap C$ at $|\alpha|\le3$.
If $h_1(\xi,\,\eta)$ and  $h_2(\xi,\,\eta)$ satisfy
the conditions $\partial^{\alpha}h_{1,2}\in L_1\cap C$
at $|\alpha|\le4$, then $h_1$ and $h_2$ may be represented in the form:
\bb
\begin{array}{c}
h_1={-1\over\pi}<\psi^{(1)}(\xi,\eta,l),\varphi^{(1)}(\xi,\eta,k)>_{\hat h},\\
h_2={1\over\pi}<\psi^{(2)}(\xi,\eta,l),\varphi^{(2)}(\xi,\eta,k)>_{\hat h},
\end{array}
\label{inverse}
\ee
where
\bb
\begin{array}{c}
\hat h_1={1\over4\pi}(\psi^+_{(1)}(\xi,\eta,k),\phi_{(1)}(\xi,\eta,l))_h,\\
\hat h_2={1\over4\pi}(\psi^-_{(2)}(\xi,\eta,k),\phi_{(2)}(\xi,\eta,l))_h.
\end{array}
\label{direct}
\ee
\end{theorem}
\par
Of course, if $q_1=q_2=0$, then the formulae (\ref{direct}) and (\ref{inverse})
are ordinary Fourier transform with respect to two variables
$\xi,\eta\in\Real$.

\section{The Fourier method for  the linearized
Davey-Ste\-wart\-son I equation.}
\par
We shall consider the Davey-Stewartson I equation:
\bb
\begin{array}{cc}
i\partial_t Q+(\partial_\xi^2+\partial_\eta^2)Q+(g_1+g_2)Q=0,\\
\partial_\xi g_1=-{\varepsilon\over2}\partial_\eta |Q|^2,\quad
\partial_\eta g_2=-{\varepsilon\over2}\partial_\xi |Q|^2,\quad
\varepsilon=\pm1.
\end{array}
\label{ds1}
\ee
\par
Linearization of this equation on $Q,\,g$ as a background gives:
\bb
\begin{array} {cc}
i\partial_t U+(\partial^2_\xi+\partial_\eta^2)U+(g_1+g_2)U+(V_1+V_2)Q=0,
\\
\partial_\xi V_1=-{\varepsilon\over2}\partial_\eta (Q\bar U+\bar Q U),\quad
\partial_\eta V_2=-{\varepsilon\over2}\partial_\xi (Q\bar U+\bar Q U).
\end{array}
\label{lds1}
\ee
The equation (\ref{ds1}) is a compatible condition for (\ref{de}) at
$q_1=Q$, $q_2=\varepsilon\bar Q$ and for a following system \cite{A-S}:
$$
\partial_t \psi=
i\Big(\begin{array}{cc} 1&0\\0&-1\end{array}\Big)
(\partial_\xi-\partial_\eta)^2\psi+
i\Big(\begin{array}{cc} 0&q_1\\q_2&0\end{array}\Big)
(\partial_\xi-\partial_\eta)\psi+
$$
\bb
+\Big(\begin{array}{cc} ig_1&-i\partial_\eta q_1\\i\partial_\xi q_2&-ig_2
\end{array}\Big)
\psi.
\label{dtpsi}
\ee
\par
Using the systems (\ref{de}) and (\ref{dtpsi}) one can prove
the following statement:
\begin{theorem}
Let $U$ be the solution of (\ref{lds1}) and $Q$ be the solution of
the Davey-Stevartson I equation with boundary conditions:
$g_1|_{\xi\to\infty}=0$ and $g_2|_{\eta\to\infty}=0$,  and
$U$ and $Q$ satisfy the conditions of the theorem 1 for the functions
$h_1$ and $q_1$ at $\forall t\in[0,T]$ respectively, then:
\bb
\partial_t \hat U_1=i(k^2+l^2)\hat U_1,\quad
\partial_t \hat U_2=-i(k^2+l^2)\hat U_2,
\label{efc}
\ee
where $\hat U_1={1\over4\pi}(\psi^{-}_{(1)},\phi_{(1)})_U$,
$\hat U_2={1\over4\pi}(\psi^{+}_{(2)},\phi_{(2)})_U$.

{\bf Inverse statement.} Let $\hat U_{1}(t,k,l)$ and $\hat U_{2}(t,k,l)$
be integrable with respect to $k,\,l$ with factor $(1+k^2)(1+l^2)$ and
satisfy equations (\ref{efc}), then the function
$$
U(\xi,\eta,t)={-1\over\pi}
<\psi^{(1)}(\xi,\eta,l),\varphi^{(1)}(\xi,\eta,k)>_{\hat U}
$$
satisfies linearized Davey-Stewartson I equation (\ref{lds1})
\end{theorem}
\par
Solving of the system (\ref{lds1}) is reduced to solving of the trivial
equations (\ref{efc}) for the Fourier coefficients $\hat U_{1,2}$.
The theorems 1 and 2 give solving (\ref{lds1}) by Fourier method.

\end{document}